\documentclass[journal]{IEEEtran} % Comment this line out if you need a4paper

\usepackage{bigints}
\usepackage{amsmath}
\newcommand{\Int}{\bigintss\limits}
\usepackage{graphicx}              

\usepackage{amsmath,amssymb}
\usepackage{graphicx}
\usepackage{float}
\usepackage{array}
\usepackage[export]{adjustbox}
\usepackage{caption}
\usepackage[justification=centering]{caption}
\usepackage{multirow}
\usepackage{booktabs}
\usepackage{hhline}
\usepackage{xcolor}
\usepackage{flushend}

\usepackage{mathtools}

\begin{document}

% \title{\LARGE \bf
% Arousal Level Estimation Using Multitaper Spectrum Estimation at Low-Power Wearable Devices}

\title{\LARGE \bf
Single-Channel EEG Based Arousal Level Estimation Using Multitaper Spectrum Estimation at Low-Power Wearable Devices}

% \title{\LARGE \bf
% Arousal Level Estimation Using 1-Channel EEG at Low-Power Wearable Devices}

\author{
Berken Utku Demirel$^1$, Ivan Skelin$^2$, Haoxin Zhang$^{2,4}$, Jack J. Lin$^{2,3,4}$, and  Mohammad Abdullah Al Faruque$^1$ \\
$^1$Department of Electrical Engineering and Computer Science,
$^2$Department of Neurology \\
$^3$Department of Anatomy and Neurobiology,
$^4$Department of Biomedical Engineering \\
University of California, Irvine, California, USA \\
\IEEEauthorblockA{{(bdemirel, iskelin, haoxinz1, linjj, alfaruqu)@uci.edu}}}
% \thanks{B.U. Demirel and M.A. Al Faruque are with the Department of Electrical Engineering and Computer Science, University of California, Irvine, CA 92697, USA, (e-mails: bdemirel@uci.edu, alfaruqu@uci.edu)}
% \thanks{I. Skelin is with the Department of Neurology, University of California, Irvine, CA 92697, USA, (e-mail: iskelin@uci.edu ).}
% \thanks{H. Zhang is with the Department of Neurology, Department of Biomedical Engineering, University of California, Irvine, CA 92697, USA, (e-mail: haoxinz1@hs.uci.edu ).}
% \thanks{J.j Lin. is with the Department of Neurology, Department of Anatomy and Neurobiology and Department of Biomedical Engineering, University of California, Irvine, CA 92697, USA, (e-mail: linjj@uci.edu).}

\maketitle
\thispagestyle{empty}
\pagestyle{empty}

\begin{abstract}
This paper proposes a novel lightweight method using the multitaper power spectrum to estimate arousal levels at wearable devices. We show that the spectral slope (1/f) of the electrophysiological power spectrum reflects the scale-free neural activity. To evaluate the proposed feature's performance, we used scalp EEG recorded during anesthesia and sleep with technician-scored Hypnogram annotations. It is shown that the proposed methodology discriminates wakefulness from reduced arousal solely based on the neurophysiological brain state with more than 80\% accuracy. Therefore, our findings describe a common electrophysiological marker that tracks reduced arousal states, which can be applied to different applications (e.g., emotion detection, driver drowsiness). Evaluation on hardware shows that the proposed methodology can be implemented for devices with a minimum RAM of 512 KB with 55 mJ average energy consumption. 

\end{abstract}


\section{Introduction and Related Work}
The assessment of the arousal level, which is defined as the level of consciousness  \cite{Arousal}, can be used for many different purposes at several applications, such as emotion recognition \cite{emotion}, sleep stage classification, or driver drowsiness. Therefore, it is important to define a common electrophysiological marker that tracks reduced arousal states, which can be employed for different applications. State-of-the-art works extract several features from the EEG signal for various classification purposes instead of finding a feature representing the underlying mechanism of neurological activity. For example, authors in \cite{feature1} extract 181 distinct frequency-domain features for classifying a person's arousal while watching emotional content videos. In \cite{feature2}, authors extract 104 features from different frequency bands of electroencephalogram (EEG) to classify sleep stages.  Also, these algorithms used more than one EEG channel to extract features. However, as the number of extracted features and channels increases, the algorithms' computational complexity and memory requirement increase, making them unsuitable for wearable devices with constrained memory and energy resources. Deep learning algorithms like Convolutional Neural Networks (CNN) are widely used \cite{DeepSleepNet, DeepSleepNet2} to avoid the feature extraction step since they automatically extract features through convolution layers. However, their memory requirements are huge due to the number of parameters in the designed architecture. For example, authors in \cite{DeepSleepNet},  use a Convolutional Neural Network (CNN) followed by Bidirectional Long Short Term Memory cells (Bi-LSTM) to classify sleep stages. The CNN layers are used for representation learning to extract time-invariant features, and the Bi-LSTM part is residual sequence learning to encode the temporal information of EEG signals. This network's major drawback is that it requires 25 epochs, where each epoch is a 30-second EEG, of data to be fed together to obtain 25 labels, which cannot be implemented in wearable devices due to memory constraints. Authors in \cite{DeepSleepNet2} proposed a  CNN architecture to classify sleep EEG to 5 different stages. The architecture needs to perform billions of multiply-accumulate operations (MACs) for classification, which is quite a time and energy consuming process. Several solutions have been proposed to make neural networks resource-efficient in terms of memory and energy, such as pruning or quantizing \cite{Nafiul, Pruning}, however, the networks' classification performance is decreased, especially when weight quantization is used as the precision of floating-operations are decreased. Moreover, these deep network architectures are designed for specific applications rather than general purposes since they do not investigate the underlying common electrophysiological markers. However, in this paper, we propose a lightweight method in terms of energy and memory to discriminate wakefulness from reduced arousals by examining the electrophysiological power spectrum of single-channel EEG.
\vspace{2mm}

The main contributions of this paper are as follows:
\begin{enumerate}
	\item A novel lightweight implementation of Multitaper power spectral estimation method to estimate arousal level in low-power wearable devices using a single channel EEG.
	
	\item Validation of our methodology on the anesthesia and sleep EEG data with technician-scored hypnogram annotations.
	
	\item Energy and memory efficiency validation of our proposed methodology on real hardware. 
	
\end{enumerate}

\section{Materials and Methods}

\subsection{Data Collection}
\label{Data}
The scalp EEG was recorded during overnight sleep or intra-operative anesthesia at the University of California at Irvine Medical Center. All patients provided informed consent according to the local ethics committees of the University of California at Berkeley and at Irvine and gave their written consent before data collection. We analyzed recordings from 8 participants between 24 and 57 years old. The Polysomnography was recorded for the sleep stages during 8 hours and 5 min quiescent rest with eyes closed before and after sleep. Data were recorded on a Grass Technologies Comet XL system (Astro-Med, Inc, West Warwick, RI) with a 19-channel EEG using the standard 10–20 setup as well as three electromyography (EMG) and four electro-oculography (EOG) electrodes are used to facilitate gold standard sleep staging. The EEG was referenced to the bilateral linked mastoids and digitized at 1000 Hz. Sleep staging was carried out by trained personnel and according to established guidelines. 
\\
The anesthesia data were recorded from the induction of anesthesia to the recovery using a Nihon Kohden recording system (256 channel amplifier), analog filtered above 0.01 Hz, and digitally sampled at 5 kHz. General anesthesia was induced intravenous with remifentanil (100 $\mu$g) and propofol (150 mg).
The awake state was defined as the time before the administration of propofol, and anesthesia was defined as the time after inducing remifentanil and propofol.
\vspace{-4mm}
\subsection{Methods}

\subsubsection{Pre-Processing}
Both sleep and anesthesia data are resampled to 200 Hz from 1000 and 5000 Hz for sleep and anesthesia respectively, using a FIR antialiasing lowpass filter. Then, a 10 order low-pass Butterworth filter is used with 50 Hz cut-off frequency for denoising. After filtering, sleep data is epoched into 30-second segments. In contrast, the anesthesia is segmented as 10-second to increase the number of epochs as anesthesia's duration (1-3 hours or less) is shorter than sleep (6-10 hours). It is observed that the feature's discrimination performance is best amongst the electrodes Fz, Pz, and Cz. For this study, we choose the Cz electrode for calculation of spectral slope and classification. 

\subsection{Feature Extraction}
After artifact removal and segmentation, the Multitaper approach based on discrete prolate Slepian sequences is used to compute the power spectral density (PSD) estimate from 0.5 Hz to 45 Hz with 0.5 Hz smoothing. The multitaper method gives better results than the periodogram for PSD estimation of EEG signals since it reduces the temporal variability and produces a consistent PSD. The multitaper method averages modified periodograms obtained using mutually orthogonal tapers (windows). Moreover, the obtained tapers are optimal in time-frequency concentration as those tapers are calculated using discrete prolate Slepian sequences.

\noindent\begin{minipage}{.5\linewidth}
\begin{equation}
    \lambda =  \frac{\Int_{-W}^{W} \bigl |X(f) \bigr| ^2 df}{\Int_{-F_s/2}^{F_s/2} \bigl |X(f) \bigr |^2 df} 
    \label{Slepian}
\end{equation}
\end{minipage}%
\begin{minipage}{.5\linewidth}
\begin{equation}
  \Int_{\displaystyle -F_s/2}^{\displaystyle F_s/2} \bigl |X(f) \bigr |^2 df < \infty
  \label{Slepian_const}
\end{equation}
\end{minipage}

\vspace{2mm}

The discrete prolate Slepian sequences (DPSS) arise from the following spectral concentration problem. The discrete Fourier transform (DTFT) $ \left(X(f) \right)$ of a finite time series $x[n]$, for which a sequence maximizes the ratio given in Equation \ref{Slepian}, subject to the constraint that the sequence has finite power (Equation \ref{Slepian_const}).

Where $F_s$ is the sampling rate of the sequence $x[n]$ and $\bigl |W \bigr | < F_s/2$. This ratio determines an index-limited sequence with the largest proportion of its energy in the band $[–W, W]$. This maximization leads to the eigenvalue problem is given in Equation \ref{eigen}.

\begin{equation}
    \sum_{m=0}^{N-1} \frac{sin(2\pi W(n-m))}{\pi (n-m)} g_k(m) = \lambda_k(N,W)g_k(n)
    \label{eigen}
\end{equation}

Where $\lambda_k$ is the eigenvalues, and $g_k(n)$ is the DPSS values that correspond to $k$th Slepian sequence.
The eigenvectors of this equation, $g_k(n)$, are the DPSS values, which are mutually orthogonal to each other. We have used 29 tapers for 30-second segments of sleep EEG and 9 tapers for 10-second anesthesia segments, so the first 29 and 9 DPSS are used for multitaper PSD estimation. After obtaining DPSS values, the modified periodograms are calculated in Equation \ref{eq:modified_peri} using a different Slepian sequence for each window.

\begin{equation}
    S_k(f) = \Delta t \biggl | \sum_{n=0}^{N-1} g_k(n) x(n) e^{-j2\pi fn\Delta t} \biggr |^2
    \label{eq:modified_peri}
\end{equation}

Here $S_k(f)$ is the modified periodograms, each obtained using a different Slepian sequence $(g_k(n))$. Finally, the multitaper PSD estimate is calculated, by averaging the modified periodograms using Equation \ref{result}.

\begin{equation}
    S(f) = \frac{1}{K} \sum_{k=0}^{K-1}S_k(f)
    \label{result}
\end{equation}

\begin{figure}[H]
    \centering
    \includegraphics[scale = 0.45]{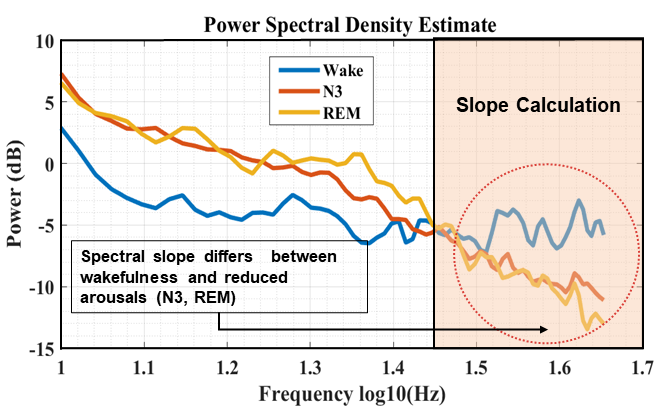}
    \caption{The power spectral density of three different EEG epochs}
    \label{fig:PSD}
\end{figure}

After obtaining the power spectral density estimation, we calculated the spectral slope by fitting a linear regression line to the PSD in log-log space between 30 and 45 Hz. This range was proved to correlate best with arousal changes in rodents and monkeys \cite{1/f}. The best line is obtained using the polynomial curve fitting method to the 30-45 Hz range of multitapered power spectral density estimation.
Figure \ref{fig:PSD} shows the PSD estimation of three different EEG epochs obtained using curve fitting.

\begin{figure}[h]
    \centering
    \includegraphics[scale = 0.4]{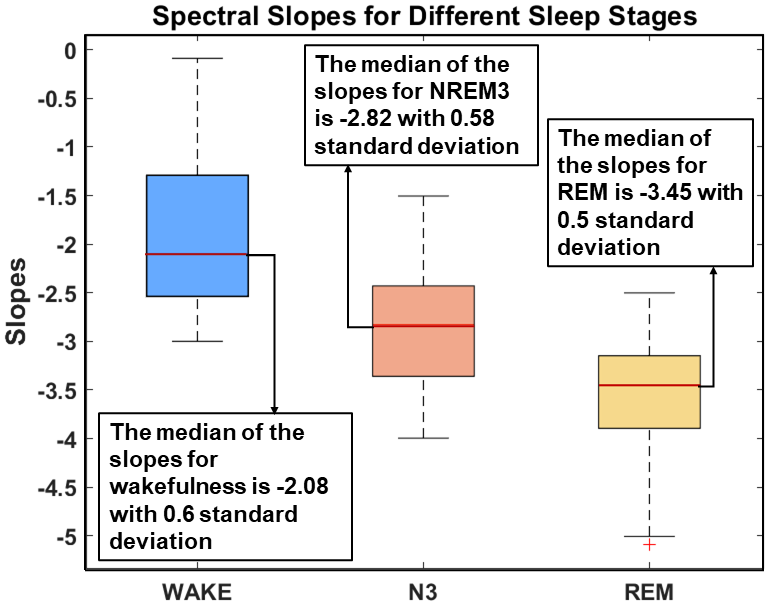}
    \caption{The whisker plot for slope distribution of three different stage}
    \label{fig:Whisker1}
\end{figure}

As shown in this figure, the spectral slope of the wakefulness and reduced arousals is different. The slope for the reduced arousals tends to be more negative than the wake stage. To observe these spectral slope differences better between sleep stages, the whiskey plot is shown in Figure \ref{fig:Whisker1}.

Figure \ref{fig:Whisker1} shows the spectral slopes' distribution for three different sleep stages (Wake, NREM3, and REM). We observed the wake stage slope between -0.1 and -3 with a -2.08 median value and 0.6 standard deviations. On the other hand, the median of spectral slopes for the REM stage is -3.45 with 0.5 standard deviations.

\begin{figure}[h]
    \centering
    \includegraphics[scale = 0.4]{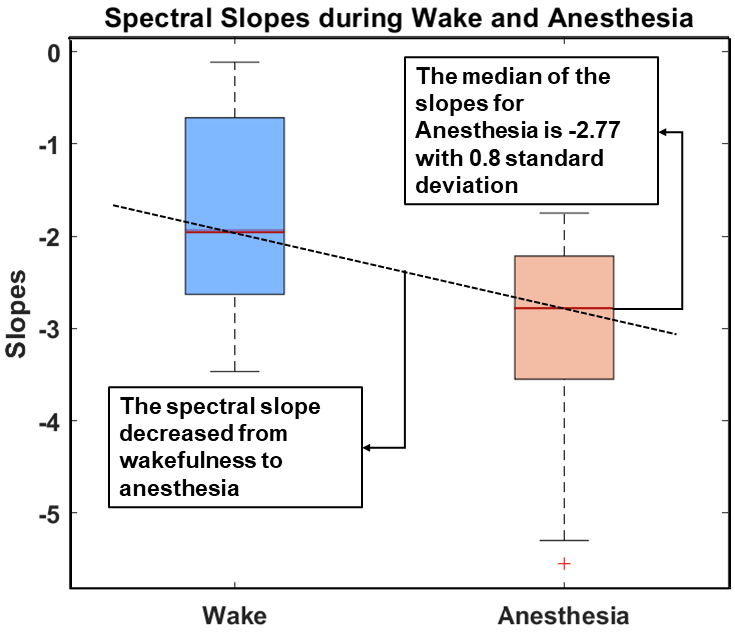}
    \caption{The whisker plot for slope distribution of wake and anesthesia}
    \label{fig:Whisker2}
\end{figure}

We have applied the same method to EEG data recorded under propofol anesthesia to show that the spectral slope between 30-45 Hz can track the reduced arousal levels. In these recordings, the wake stage is defined as the time before the start of propofol and remifentanil, and anesthesia was defined as the periods when the patients were unresponsive to verbal commands assessed by the attending anesthetist.
Figure \ref{fig:Whisker2} shows the spectral slopes obtained during wakefulness and anesthesia. We observed that the spectral slope was higher during wakefulness (-1.9 median value with 0.5 standard deviations) than during anesthesia (-2.77 median value with 0.8 standard deviations). 

These results provide evidence that the spectral slope discriminates wakefulness from reduced arousals. Moreover, it is observed that the more negative slopes of the wake stage are closer to state transitions, which can be explained as the subjects are still drowsy. 

\section{Results and Discussion}

\subsection{Performance Evaluation}
To evaluate the proposed methodology's performance, we used sleep recording, specified in Section \ref{Data}, which has an associated hypnogram file scored by a specialist. The hypnogram files contain labels identifying the sleep stages. The two different NREM stages (3 and 4) are combined into one group as NREM3 since these two stages are considered as deep sleep, whereas NREM1 and NREM2 are considered as light sleep \cite{LightSleep}. Thus, the classification is performed for three different sleep stages (Wake, NREM3, and REM).

Since we extracted spectral slope from one channel of the EEG signal, simple threshold values are used for discriminating sleep stages from each other instead of using machine learning algorithms.  For this study, the threshold value is chosen as -2.45 for separating wakefulness from different sleep stages. If the calculated spectral slope of an EEG epoch is greater than -2.45, this epoch is classified as \textit{Wake}. If the slope is less than -3.2, the period is classified as \textit{REM}. Lastly, if the calculated spectral slope is between these two threshold values, the epoch is classified as the \textit{NREM3} stage.

\begin{figure}[h]
    \centering
    \includegraphics[scale = 0.5]{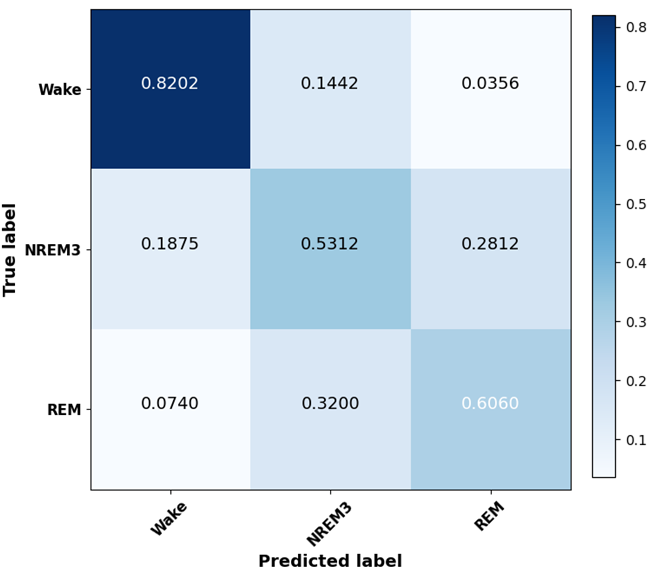}
    \caption{The normalized confusion matrix }
    \label{fig:confuse}
\end{figure}

Figure \ref{fig:confuse} shows the confusion matrix for three different sleep stages. As can be seen from the figure, the confusion of \textit{Wake} stage with \textit{NREM3} and \textit{REM} stages are lower than $15\%$, which indicates that the spectral slope feature can discriminate the wakefulness with reduced arousal brain states. Since both \textit{NREM3} and \textit{REM} stages are considered as low arousal level, the algorithm's discrimination performance for these two stages is not as high compared to \textit{Wake} classification. However, the confusion degree for reduced arousal with wakefulness is as low as $7\%$ (Wake-REM). As seen in these results, the spectral slope calculated from the power spectral density can estimate the human's arousal level based on neurophysiological brain state. Also, it is known that if the PSD estimates are obtained from different electrodes with high sampling frequency and averaged across all electrodes, the discrimination performance of the spectral slope increases for arousal level estimation \cite{janna} or different methods and techniques can be applied to multitaper power spectra to obtain a more accurate and fine estimation of frequency variations.  However, the algorithm's memory requirement and energy consumption increase with the sampling rate and number of channels used, making it unsuitable for wearable devices. Also, most of the wearable system try to use as few  channels as possible to increase the comfortability and acceptability amongst society \cite{neurotechnology-aicps}.  

\vspace{-2mm}

\subsection{Memory and Energy Consumption Evaluation}
We evaluate our proposed methodology's memory footprint and energy consumption using an STM32 Nucleo-144 (STM32H743Z), 32-bit ARM Cortex-M7 core with 480 MHz maximum operating frequency, and 1 MB RAM. The energy profiling is performed using STM32 Power Shield (LPM01A), an accurate power monitoring device with a power consumption measurement range between 180 nW and 165 mW. Table  \ref{tab:energy} shows the execution time, energy consumption,  and required memory for each operation. The energy and memory calculation is performed using a 30-second epoch of EEG signal, sampled at 200 Hz.

% \begin{table}
% \centering
% \caption{Memory and Energy Consumption on Nucleo Board}
% \label{tab:energy}
% \begin{adjustbox}{width=\columnwidth,center}
% \begin{tabular}{|c|c|c|c|c|} 
% \hline
%  \textbf{Operations}  & \begin{tabular}[c]{@{}c@{}}\textbf{Exe.}\\\textbf{Time (ms)} \end{tabular} & \begin{tabular}[c]{@{}c@{}}\textbf{Energy}\\\textbf{mJ} \end{tabular} & \begin{tabular}[c]{@{}c@{}}\textbf{Flash Memory}\\\textbf{Footprint (KB)} \end{tabular} & \begin{tabular}[c]{@{}c@{}}\textbf{RAM}\\\textbf{Footprint (KB)} \end{tabular} \\ 
% \hhline{|=====|}
% \begin{tabular}[c]{@{}c@{}}\textbf{Butterrworth}\\\textbf{Filtering} \end{tabular} & 180 & 34 & 26.4 & 96.87 \\ 
% \hline
% \begin{tabular}[c]{@{}c@{}}\textbf{Multitaper PSD}\\\textbf{Estimation} \end{tabular} & 65 & 17.5 & 45 & 350 \\
% \hline
% \end{tabular}
% \end{adjustbox}
% \end{table}

\begin{table}[H]
\centering
\caption{Memory and Energy Consumption on Nucleo Board}
\label{tab:energy}
\begin{adjustbox}{width=\columnwidth,center}
\begin{tabular}{|c|c|c|c|c|} 
\hline
 \textbf{Operations}  & \begin{tabular}[c]{@{}c@{}}\textbf{Exe.}\\\textbf{Time (ms)} \end{tabular} & \begin{tabular}[c]{@{}c@{}}\textbf{Energy}\\\textbf{mJ} \end{tabular} & \begin{tabular}[c]{@{}c@{}}\textbf{Flash Memory}\\\textbf{Footprint (KB)} \end{tabular} & \begin{tabular}[c]{@{}c@{}}\textbf{RAM}\\\textbf{Footprint (KB)} \end{tabular} \\ 
\hhline{|=====|}
\textbf{Filtering} & 180 & 10 & 26.4 & 96.87 \\ 
\hline
\begin{tabular}[c]{@{}c@{}}\textbf{Multitaper PSD}\\\textbf{Estimation} \end{tabular} & 101 & 45 & 45 & 350 \\
\hline
\end{tabular}
\end{adjustbox}
\end{table}

The overall execution time for a 30-second epoch takes 281 ms in the device with 55 mJ average energy consumption. The proposed methodology is also compatible with any devices with a minimum RAM of 512 KB. When multitaper PSD estimation is calculated, the discrete Prolepian sequences are represented as sparse matrices to save memory. Since the first sequences contain a few non-zero elements, storing them as sparse matrices are more efficient. The number of discrete Slepian sequences (29 tapers in this study) used for PSD calculation can be decreased using fewer tapers to save more memory. However, the value for frequency smoothing needs to be decreased, which would reduce PSD resolution and result in a high-variance spectrum.
\vspace{1mm}

\section{Conclusions}
This paper proposes a lightweight feature extracted using Multitaper power spectra that can be implemented in low-power wearable devices to discriminate the reduced arousal states from wakefulness based on single-channel EEG signals. We used EEG signals recorded in two different ways during sleep and anesthesia to validate the performance of the feature. It is shown that the feature is an electrophysiological marker that tracks reduced arousal states in EEG signals. Therefore, it can be applied for various applications in which monitoring the arousal level is important such as emotion recognition, epileptic seizures, coma, the vegetative or minimally conscious state. The implementation of the proposed methodology on hardware shows that it is energy and memory-efficient with 55 mJ average energy consumption, making it suitable for wearable devices with a minumum RAM of 512 KB.


\bibliographystyle{IEEEtran} % We choose the "plain" reference style
\bibliography{ref}

\end{document}